\documentclass[12pt,preprint]{aastex}

\slugcomment{Not to appear in Nonlearned J., 45.}

\shorttitle{AGN-Host Connection in Partially Obscured AGNs. I}
\shortauthors{Wang \& Wei}

\begin{document}


\title{UNDERSTANDING AGN-HOST CONNECTION IN PARTIALLY OBSCURED ACTIVE GALACTIC NUCLEI.\\
Part III: PROPERTIES OF ROSAT-SELECTED SDSS AGNs \\ }


\author{J. Wang and J. Y. Wei}
\affil{National Astronomical Observatories, Chinese Academy of Science, 20A Datun Road, 
Chaoyang District, Beijing 100012, China}

\email{wj@bao.ac.cn}


%\altaffiltext{1}{Visiting Astronomer, Cerro Tololo Inter-American Observatory.
%CTIO is operated by AURA, Inc.\ under contract to the National Science
%\altaffiltext{2}{Society of Fellows, Harvard University.}
%\altaffiltext{3}{present address: Center for Astrophysics,
%    60 Garden Street, Cambridge, MA 02138}
%\altaffiltext{4}{Visiting Programmer, Space Telescope Science Institute}
%\altaffiltext{5}{Patron, Alonso's Bar and Grill}


\begin{abstract}
As the third paper of our serial studies that are aim at examining the AGN-host coevolution by 
using partially obscured AGNs, we extend the broad-line composite galaxies (composite AGNs) into 
\it ROSAT\rm-selected Seyfert 1.8/1.9
galaxies basing upon the RASS/SDSS-DR5 catalog given by Anderson et al.. 
The SDSS spectra of in total 92 objects are analyzed by the same method used in our previous studies, 
after requiring the signal-to-noise ratio in the SDSS $r'$ band is larger than 20. 
Combing the \it ROSAT\rm-selected Seyfert galaxies with the composite AGNs
reinforces the tight correlation 
between the line ratio [\ion{O}{1}]/H$\alpha$ vs. $D_n(4000)$, and establishes a new tight correlation 
between [\ion{S}{2}]/H$\alpha$ vs. $D_n(4000)$. Both correlations suggest the two line ratios are 
plausible age indicators of the circumnuclear stellar population for typical type I AGNs in which the 
stellar populations are difficult to be derived from their optical spectra. The \it ROSAT\rm-selected Seyfert galaxies
show that the two correlations depend on the soft X-ray spectral slope $\alpha_X$ that 
is roughly estimated from the hardness ratios by requiring the X-ray count rates within 0.1-2.4 keV are larger 
than $0.02\ \mathrm{counts\ s^{-1}}$. However, we fail to establish a relationship between 
$\alpha_X$ and $D_n(4000)$, which is likely caused by the relatively large uncertainties of both 
parameters (especially for $\alpha_X$ because of the AGN intrinsic obscuration).
The previously established $L/L_{\mathrm{Edd}}-D_n(4000)$ evolutionary sequence is 
reinforced again by the extension to the \it ROSAT\rm-selected Seyfert galaxies. These 
X-ray-selected Seyfert galaxies are, however, biased against the two ends of the sequence, which implies that 
the X-ray Seyfert galaxies present a population at middle evolutionary stage.

\end{abstract}


\keywords{galaxies: active --- galaxies: nuclei ---  galaxies: evolution}



\section{INTRODUCTION}

Active Galactic Nuclei (AGNs) are now widely believed to co-evolve with
their host galaxies. The co-evolution scenario is based upon the two observational facts.  
One is the firmly established tight Magorrian relationship (i.e., the 
$M_{\mathrm{BH}}-\sigma_*$ or $M_{\mathrm{BH}}-L_{\mathrm{bugle}}$ relationship, 
e.g., Magorrian et al. 1998; Tremaine et al. 2002; Ferrarese \& Merritt 2000; Greene \& Ho 2006;
Greene et al. 2008; Haring \& Rix 2004). Another one is the fact that the global evolutionary 
history of growth of central supermassive black hole (SMBH) and star formation history trace each other closely 
from present to redshift $z\sim5$ (e.g., Nandra et al. 2005; Silverman et al. 2008; 
Shankar et al. 2009; Hasinger et al. 2005). A lot of efforts in both theoretical and observational grounds
have been made to understand the elusive 
evolutionary connection between AGNs and their host galaxies over the past decades.
For example, young stellar populations and enhanced ongoing star formation activities are 
prevalently identified in host galaxies of many Seyfert galaxies and quasars 
(e.g., Gonzalez Delgado 2002; Silverman et al. 2009; White \& Nelson 2003; Zuther et al. 2007;
Canalizo \& Stockton 2001; Stockton et al. 2007; Riffel et al. 2008; Cid Fernandes et al. 2001, 2005;
Davies et al. 2007; Canalizo et al. 2007; Wang \& Wei 2006; Zhou et al. 2005; Mao et al. 2009;
Heckman \& Kauffmann 2006). In addition, many authors established a relationship between the black 
hole accretion 
properties (e.g., $L/L_{\mathrm{Edd}}$) and properties (e.g., age and star formation rate) of 
stellar population of bulge of AGN host 
galaxies (e.g., Kewley et al. 2006; Wild et al. 2007; Wang et al. 2006; Wang \& Wei 2008; Watabe et al. 2008; 
Davies et al. 2007). Theoretical simulations indicated that a major merger between two 
gas-rich disk galaxies is a plausible model for the co-evolution (e.g., Di Matto et al. 2007; 
Hopkins et al. 2005, 2007; Granato et al. 2004). 
Finally, there are accumulating evidence supporting that the detectable
AGN activity delays for $\sim10^2$ Myr after the onset of the star formation activity 
(e.g., Hopkins et al. 2005; Wang \& Wei 2006; Wang et al. 2009b; Schawinski et al. 2009; 
Reichard et al. 2008; Li et al. 2007).
  
Nevertheless, the details of the co-evolution issue are still far from fully understood at the 
current stage. The main difficulty in understanding the coevolution
is the orientation effect caused by the AGN's torus (see reviews in Antonucci 1993;
Elitzur 2007). We refer the readers to Wang \& Wei (2008, and references therein, hereafter Paper I) for 
a brief comment on the several approaches that were adopted to overcome the difficulty.
In Paper I, the spectra of partially obscured AGNs were used to simultaneously 
measure the properties of both AGN and stellar population in an individual object.
By examining the spectral properties of the 85 composite AGNs\footnote{These composite AGNs 
with broad H$\alpha$ emission lines are selected from the composite galaxies that are 
located in the [\ion{O}{3}]/H$\beta$ vs. [\ion{N}{2}]/H$\alpha$ diagnostic diagram between the empirical and theoretical 
demarcation lines that separate AGNs from starforming galaxies.} selected from 
the SDSS DR4 MPA/JHU catalog (see Heckman \& Kauffmann 2006 for a review),
we identified a $L/L_{\mathrm{Edd}}-D_n(4000)$ evolutionary sequence and a tight 
correlation between the spectral break
at 4000\AA\ ($D_n(4000)$) and line flux ratio [\ion{O}{1}]$\lambda6300$/H$\alpha$.

The study in Paper I is, however, far from a complete understanding of the co-evolution issue.
At first, Paper I only focused on the composite galaxies whose narrow emission lines are 
believed to contain significant contributions from both star formation and AGN.
The stellar populations in the composite galaxies are found to be systematically young
(e.g., Kewley et al. 2006; Wang \& Wei 2008; Schawinski et al. 2009), which
naturally causes the composite AGNs studied in Paper tend to cluster around the small
$D_n(4000)$ end in the two relationships.
Basing upon the narrow emission-line galaxies from the MPA/JHU catalog, Kewley et al. (2006) suggested 
that the composite galaxies might evolve to Seyfert galaxies\footnote{The Seyfert galaxies
are defined as the emission-line galaxies whose location in the [\ion{O}{3}]/H$\beta$ vs. [\ion{N}{2}]/H$\alpha$
diagram is above the theoretical demarcation lines that
separate AGNs from starforming galaxies (see Section 3.2 for details).} (see also in Constantin et al. 2009).
An inclusion of the Seyfert galaxies (and also LINERs) is therefore 
necessary for obtaining a complete understanding of the co-evolution issue.

Secondly, the X-ray emission of AGNs that are not examined in Paper I 
might play an important role in the co-evolution.  
The theoretical calculations suggested that the line ratio [\ion{O}{1}]/H$\alpha$
is sensitive to the hardness of the ionizing field (Kewley et al. 2006). The tight 
correlation between [\ion{O}{1}]/H$\alpha$ and $D_n(4000)$ therefore  
implies an evolutionary significance of X-ray emission of AGNs.
The evolutionary role of X-ray emission is also indirectly suggested by the
study of the Eigenvector I (EI) space of AGNs.
The EI space was first introduced by Boroson \& Green (1992)
who analyzed the optical spectra of 87 bright PG quasars, and was subsequently extended to
soft X-ray spectral slope $\Gamma_\mathrm{soft}$
(e.g., Wang et al. 1996; Grupe 2004; Xu et al. 2003, and see
Sulentic et al. 2000 for a review). 
By examining the optical spectra of a sample of
\it IRAS\rm-selected Seyfert 1.5 galaxies, Wang et al. (2006) first extended the EI space into
middle-far infrared colors $\alpha(60,25)$, which implies a relation
between the essential EI space and age of circumnuclear stellar population.

A sample of X-ray selected Seyfert galaxies is therefore not only appropriate to the extension of the 
two relationships (i.e., the $L/L_{\mathrm{Edd}}-D_n(4000)$ sequence and 
[\ion{O}{1}]/H$\alpha$ vs. $D_n(4000)$ correlation) 
to Seyfert galaxies, but also required by the study of the speculated evolutionary role of X-ray emission. 
It is widely accepted that luminous X-ray emission is a common property for typical AGNs.
Typically, X-ray luminosity in AGNs contributes a fraction of about 5-40\% bolometric luminosity.
A large fraction of the counterparts of the X-ray sources are optically spectroscopically
identified as AGNs (e.g., Stocke et al. 1991; Obric et al. 2006; Anderson et al. 2007;
Kollatschny et al. 2008; Mahoney et al. 2010; Wei et al. 1999).
Obric et al. (2006) found the number of AGNs is six times more than starforming galaxies for
SDSS-ROSAT narrow emission-line galaxies according to the [\ion{O}{3}]/H$\beta$ vs. [\ion{N}{2}]/H$\alpha$
diagnostic diagram.
Because the properties of stellar populations are required in our study,
we focus on partially obscured AGNs again by following the same approach used in Paper I 
since their spectra have an advantage of balance of contribution of AGN and starlight.

With these motivations, we extend the study in Paper I to soft X-ray selected partially obscured Seyfert
galaxies by selecting these galaxies from the combination of 
the SDSS DR5 spectroscopic survey and \it ROSAT\rm\ All Sky Survey 
(hereafter \it ROSAT \rm -selected Seyfert galaxies). The cross-match was originally 
done by Anderson et al. (2007). The selected sample is combined with 
the composite AGNs studied in Paper I to provide a complete understanding of the co-evolutionary issue, except for the 
analysis in which the X-ray properties are required.  
The paper is structured as follows. \S2 describes the sample selection and data reduction. The 
physical properties of both AGNs and stellar populations are obtained in \S3. \S4 and \S5 present 
our analysis and discussion, respectively. The $\Lambda$ cold dark matter ($\Lambda$CDM) cosmology
with parameter $h_0=0.7$, $\Omega_{\rm M}=0.3$, and $\Omega_{\Lambda}=0.7$ (Spergel et al. 2003)
is adopted throughout the paper.

\section{SAMPLE AND DATA REDUCTION}

\subsection{The Sample of SDSS DR5/RASS Seyfert 1.8/1.9 Galaxies}

Anderson et al. (2007 and references therein) recently carried out a RASS/SDSS program 
to identify the optical counter parts of the \it ROSAT\rm\ X-ray sources from the SDSS 
Data Release 5 catalog, simply because the depth of the two surveys are well matched. 
The updated catalog contains about 7,000 quasars/broad-line AGNs and 500 narrow-line AGNs 
(including Seyfert 1.5-1.9 galaxies) with plausible X-ray emission.
These AGNs were classified by Anderson et al. (2007) through visual examination of their 
SDSS spectra. We only focus on the Seyfert 1.8/1.9 galaxies in this paper. 
Among these  
Seyfert 1.8/1.9 galaxies, we further require the redshifts range from 0.03 to 0.3, which 
is comparable with the study in Paper I. In order to ensure that our results are 
not affected by the spectra with poor quality, the objects with  
signal-to-noise ratio in $r'$ band image larger than 20 are only considered in our subsequent
analysis. There are finally a total of 93 \it ROSAT\rm-selected Seyfert galaxies that are listed 
in our final sample. 

RASS surveyed nearly all sky down to
a limiting sensitivity $\sim10^{-13}\ \mathrm{ergs\ s^{-1}\ cm^{-2}}$. 
The sensitivity limit therefore results in a selection effect that 
leads us to miss the AGNs with low X-ray emission (both intrinsic and obscured, 
see Section 4.2 for details).

The position error circles are about 10\symbol{125}-30\symbol{125} for RASS. The probability of true match 
between SDSS and RASS strongly depends not only on the angular distance between SDSS main galaxy sample 
and RASS sources, but also on the spectral types of galaxies. The simulation in Parejko et al. (2008) 
indicates that a) the probability of 
a reliable cross-match between SDSS galaxies and RASS sources is larger than $\sim$40\% when the seperation 
is within $\sim30$\symbol{125}; b) this probability is greatly enhanced to $\sim95$\% if the galaxies 
can be classified as AGNs with broad emission lines (see also Anderson et al. 2007); c) 
a X-ray counterpart could be always reliably identified for LINERs,  Seyfert 2s,
transitions and unclassified galaxies, except for starforming galaxies. The cross-match 
radius adopted in Anderson et al. (2007) is 27.5\symbol{125} away from each of the bright X-ray
source. Our spectral analysis shows that 
none of the \it ROSAT-\rm selected galaxies used in the current study is 
classified as a starforming galaxy. The redshifts are displayed per spectral type in Figure 1 for the 
\it ROSAT-\rm selected Seyfert galaxies. The mean and median values are marked by the open triangle 
and square, respectively, for each of the spectral type. We classified these galaxies as transitions (T), Seyfert 
galaxies (S), LINERs (L) and unclassified galaxies (U) following the recent classification 
scheme proposed in Kewley et al. (2006) using the Baldwin-Phillips-Terlevich (BPT) diagrams (see Section 3.2). 
All four types of 
galaxy show similar both mean and median redshifts at around $z\sim0.1$.

\subsection{Spectral Analysis}

The SDSS spectra are reduced by following the identical method that was described in 
detail in Paper I. Briefly,
the Galactic extinction is first corrected for each spectrum by using the color excess 
$E(B-V)$ taken from the Schlegel, Finkbeiner and Davis Galactic reddening map (Schlegel et al. 1998).
The extinction law with $R_V=3.1$ is adopted for the correction (Cardelli et al. 1989). 
Each extinction-corrected spectrum is then transformed to the rest frame, along with the 
$k$-correction, given the redshift measured by the SDSS pipelines (Glazebrook et al. 1998; 
Bromley et al. 1998).

Among the whole sample, the continuum of a majority of the spectra (90\%=83 objects) is dominated 
by the stellar absorption features. 
For each of these objects, a pipeline based on the 
principal component analysis (PCA) method has been developed by us to separate the 
starlight from the observed spectrum. We refer the readers to Paper I for 
the details of the PCA method. The modeling and removing of the starlight component is illustrated 
for one typical object 
in the upper panel of the left column of Figure 2. However, the starlight component can not be 
correctly separated from the observed spectra for the remaining 13 objects, because their continuum
is dominated by the featureless emission from central AGNs. In these cases, the 
continuum is instead modeled by a powerlaw through a $\chi^2$ minimization. The minimization is carried out 
in the rest-frame wavelength ranging from 4000\AA\ to 7000\AA, except for the regions around the 
strong AGN emission lines: e.g., Balmer lines, [\ion{O}{3}]$\lambda\lambda$4959, 5007, 
\ion{He}{2}$\lambda$4686, [\ion{N}{2}]$\lambda\lambda$6583, 6548, [\ion{S}{2}]$\lambda\lambda$6716, 6731,
[\ion{O}{3}]$\lambda$4363, and possible \ion{Fe}{2} complex from 4200\AA\ to 4500\AA. In fact, 
the \ion{Fe}{2} blends are not included in the continuum modeling and in the subsequent analysis,
because the \ion{Fe}{2} blends are too weak to be detectable in these spectra. The lower panel 
in the left column in Figure 2 shows an illustration for the powerlaw continuum modeling.

The emission lines are modeled and measured by the 
SPECFIT task (Kriss 1994) in the IRAF package\footnote{IRAF is distributed by the National 
Optical Astronomical Observatories,
which is operated by the Association of Universities for Research in Astronomy, Inc.,
under cooperative agreement with the National Science Foundation.} at the H$\beta$ and H$\alpha$ regions,
after the starlight/powerlaw continuum is removed from the observed spectra.
In each spectrum, the observed profile of each emission line is modeled by a set of Gaussian components
through their linear combination. 
The emission-line modelings in the H$\beta$ and H$\alpha$ regions are schemed in the 
middle and right columns in Figure 2, respectively. 
The flux of [\ion{O}{1}]$\lambda$6300 emission line
is measured by a direct integration in each spectrum through the SPLOT task in the IRAF. 
Although the broad H$\alpha$ emission can be usually
well modeled by a single Gaussian component in most objects, an additional very broad 
H$\alpha$ (with FWHM$\sim10000\ \mathrm{km\ s^{-1}}$) is required to adequately model 
the broad H$\alpha$ emission in five objects 
(see Paper I and references therein, and recently in Zhu et al. 2008).
We determine the FWHM of the broad H$\alpha$ emission in these particular cases as follows.
We first model the observed H$\alpha$ profile by a set of Gaussian 
components including the additional very broad component. A residual profile is then
produced by subtracting the modeled H$\alpha$ narrow component from the observed 
profile. The line width of the broad H$\alpha$ emission is finally measured on the 
residual profile through the SPLOT task. The starlight around H$\beta$ is usually 
over-subtracted, especially for the objects associated with old stellar populations, 
which results in a 2\%-7\% underestimate in the H$\beta$ flux (see Paper I and 
Asari et al. 2007). The over-subtraction is probably caused by the calibration in 
the STELIB library (Asari et al. 2007). In this paper, we slightly enhance the continuum 
level around the H$\beta$ emission line from zero by acquiring all the H$\beta$ 
emission is above the zero level to account for the small underestimation. 
Although the small underestimation is corrected here, the results given in this paper
are still comparable with our study in Paper I where the small underestimation 
is ignored, both because the correction is too small to affect the main results 
and because the composite AGNs are mainly associated with relatively 
young stellar populations.

The results of the emission-line measurements are
tabulated in Table 1 and 2 for the objects whose continuum is dominated by 
starlight and by powerlaw emission, respectively. 
For each object in Table 1, Column (1) lists the SDSS identification,
and Column (2) the corresponding redshift. The line ratios in logarithm of [\ion{N}{2}]/H$\alpha$, 
[\ion{S}{2}]/H$\alpha$, [\ion{O}{1}]/H$\alpha$ and [\ion{O}{3}]/H$\beta$
are listed from Column (3) to Column (6). Column (7) and (8) show the 
FWHM of the H$\alpha$ broad component and intrinsic luminosity 
($L_{\mathrm{H\alpha}}$), respectively. The luminosity is corrected for the local extinction. 
The extinction is inferred from the narrow-line ratio H$\alpha$/H$\beta$ for each object, assuming the
Balmer decrement for standard case B recombination and Galactic extinction curve with $R_V=3.1$.
For the objects whose continuum is modeled by a powerlaw instead of starlight, Table 2 only lists
their line ratios of [\ion{N}{2}]/H$\alpha$,
[\ion{S}{2}]/H$\alpha$, [\ion{O}{1}]/H$\alpha$ and [\ion{O}{3}]/H$\beta$, since the broad H$\alpha$
components will not be used in our subsequent analysis.

\section{ANALYSIS AND RESULTS}

Basing upon the spectral measurements given above, the multiwavelength properties of the 
\it ROSAT\rm-selected Seyfert galaxies are derived and analyzed in this section. The \it ROSAT\rm-selected
galaxies are then combined and compared with the composite AGNs previously studied in Paper I to give
us a complete understanding of the coevolution of AGNs and their host galaxies.

\subsection{Deriving Physical Properties}

The black hole mass ($M_{\mathrm{BH}}$), Eddington ratio ($L/L_{\mathrm{Edd}}$), soft X-ray 
spectral slope and spectral 
indices being sensitive to age of stellar 
population are calculated basing upon the spectral analysis mentioned above.
The calculation is briefly described as follows, 
because some methods that are adopted here are the same as that 
used in Paper I. We refer the readers to Paper I for the details of the 
calculation and discussion about the uncertainties. 

\subsubsection{$L/L_{\mathrm{Edd}}$ and $M_{\mathrm{BH}}$}

Two arguments allow us to believe that 
the detected emission of the H$\alpha$ broad components is mainly contributed from central AGNs.  
At first, the detectable X-ray emission strongly supports the presence of an AGN in the center 
of the galaxies. Moreover, the minimal value of the measured luminosity of the broad H$\alpha$ 
components is $1.59\times10^{40}\ \mathrm{ergs\ s^{-1}}$. The average and median values are
$6.32\times10^{41}\ \mathrm{ergs\ s^{-1}}$ and 
$3.32\times10^{41}\ \mathrm{ergs\ s^{-1}}$, respectively, which is far larger than the 
values that could be produced by the other mechanisms (e.g., WRs and luminous blue variables 
with $L_{\mathrm{H\alpha B}}\sim10^{36-40} \mathrm{ergs\ s^{-1}}$, 
see Izotov et al. 2007; Izotov \& Thuan 2008 
and references therein; and discussions in Paper I). 
 
Following the study in Paper I, the two basic parameters of black hole accretion 
(i.e., $L/L_{\mathrm{Edd}}$ and $M_{\mathrm{BH}}$) are 
directly estimated from the AGN broad H$\alpha$ emission lines, because the continuum of these objects 
is highly contaminated by the emission from their host galaxies.
Basing upon the revised luminosity-radius relation provided by 
Bentz et al. (2009), Greene \& Ho (2007) derived an updated estimator of mass of central SMBH.
In order to compare the results given in the current paper with that reported in Paper I, we estimate 
$L/L_{\mathrm{Edd}}$ and $M_{\mathrm{BH}}$ by using the same equations used in Paper I, i.e., the 
calibrations provided in Greene \& Ho (2005).  The estimated values of $L/L_{\rm Edd}$ and $M_{\mathrm{BH}}$ 
are listed in the Column (11) and (12) in Table 1 for each object, respectively.
$L/L_{\mathrm{Edd}}$ and $M_{\mathrm{BH}}$ are not available for 14 objects in which the very weak  
broad H$\alpha$ components prevent us from estimating the values. 
As stated in Paper I, the estimated   
$L/L_{\mathrm{Edd}}$ and $M_{\mathrm{BH}}$ are likely systemically underestimated in the partially 
obscured AGNs. The upper limit of the underestimation is roughly $\sim70\%$ for $M_{\mathrm{BH}}$ and 
$\sim50\%$ for $L/L_{\mathrm{Edd}}$.

\subsubsection{$D_n(4000)$ and H$\delta_A$}

As done in Paper I, we use the 4000\AA\ break ($\rm D_n(4000)$) and equivalent width of
H$\delta$ absorption of A-type stars (H$\delta_A$) as indicators of the ages of 
stellar populations of AGN host galaxies (e.g., Heckman et al. 2004; Kauffmann et al. 2003a; 
Kewley et al. 2006). The two indices are measured in the removed starlight spectrum for each object. 
The measured values of both indices are tabulated in Column (9) and (10) in Table 1. Figure 3 compares 
the number distribution of the measured $\rm D_n(4000)$ of the \it ROSAT\rm-selected Seyfert galaxies 
with that of the composite AGNs. 
The stellar populations of the \it ROSAT\rm-selected Seyfert galaxies are systematically  
older than that of the composite AGNs, which agrees with the previous studies (e.g., Kewley et al. 2006). 
The Gehan's generalized Wilcoxon two-sample statistical test shows the two distributions 
are drawn from the same parent population at the confidence level of 1.2\%. This conclusion is further 
confirmed by a two-side Kolmogorov-Smirnov test that yields a maximum discrepancy of 0.21 with a corresponding 
probability 5.6\% that the two distributions match.

\subsubsection{Soft X-ray spectral slope $\alpha_X$}

The X-ray spectral properties are estimated from the \it ROSAT\rm\ hardness ratios HR1 and HR2\footnote{
The \it ROSAT\rm\ hardness ratios HR1 and HR2 are defined as $\mathrm{HR1}=(B-A)/(B+A)$ and 
$\mathrm{HR2}=(D-C)/(D+C)$, where $A$, $B$, $C$, and $D$ are the count rates in the energy bands 
0.1-0.4, 0.5-2.0,0.5-0.9 and 0.9-2.0 keV, respectively.} (Voges et al. 1999)
by assuming a single powerlaw (i.e., $F_\nu\propto\nu^{-\alpha_X}$)
modified by the foreground absorption resulted from our own galaxy only (e.g., Xu et al. 2003; Yuan et al. 2008;
Grupe et al. 1998). The Galactic hydrogen column density $N_{\mathrm{H}}$ is transformed from the
color excess $E(B-V)$ through the calibration $N_{\mathrm{H}}=(5.8\pm2.5)\times10^{21}E(B-V)\ \mathrm{cm^{-2}}$
(Bohlin et al. 1978) for each object, where $E(B-V)$ is adopted from the
Schlegel, Finkbeiner and Davis Galactic reddening map (Schlegel et al. 1998) again. 
The soft X-ray opacity is calculated from the effective absorption cross sections given in 
Morrison \& McCammon (1983) by assuming a solar abundance.

Previous studies 
suggested that this method is reliable and robust if the intrinsic X-ray spectra can indeed 
be described as a simple powerlaw (e.g., Brinkmann \& Siebert 1994; Schartel et al. 1996). 
The X-ray spectral slope $\alpha_X$ is estimated for the objects whose X-ray count rates
within the 0.1-2.4 keV band are larger than $0.02\ \mathrm{counts\ s^{-1}}$. 
We list the estimated $\alpha_X$ in Column (13) in Table 1 and Column (7) in Table 2. 
The estimated 
$\alpha_X$ ranges from 0.1 to 5, which statistically agrees with the previous studies that focused on 
typical Type I AGNs (e.g., Xu et al. 2003; Grupe 2004). 
Although it is the best approach for a large AGN sample at present,   
strictly speaking, local absorption (and obscuration) from AGNs and their host galaxies should be involved in
the spectral model for these partially obscured AGNs, because X-ray and optical absorptions are 
statistically correlated in AGNs (e.g., Garcet et al. 2007; Silverman et al. 2005). In fact, one can see 
from below that such simple approach greatly degrades our results.    

\subsection{BPT Diagnostic Diagrams}

The refined BPT diagrams (Baldwin et al. 1981; Vellieux \& Osterbrock 1987)
are commonly used to diagnose the 
central energy sources in narrow-line emission galaxies (see recent review in Groves et al. 2006).
According to the classification scheme recently suggested in Kewley et al. (2006),
we classified the 80 \it ROSAT-\rm selected Seyfert galaxies for which all the four line 
ratios are measured.
There are totally 13 transitions, 38 Seyfert galaxies, 1 LINERs, and 28 unclassified galaxies. 
The unclassified galaxies means they are differently classified in the three diagrams.

Recent studies indicated that the BPT diagrams reflect not only the dominant central 
energy sources, but also an evolutionary sequence of AGNs. Using the large spectral database 
of narrow-line emission galaxies selected from the SDSS, Kewley et al. (2006) found that the 
properties (e.g., $D_n(4000)$) of AGN host galaxies changes with the distance from the starforming 
sequence. Wang et al. (2009b) recently pointed out that 
the traditionally classified LINERs are consist of two populations likely at different evolutionary 
stages, with different power sources (see also in Stasinska et al. 2008).

The distributions on the three traditional BPT diagrams are shown in Figure 4 and 6 for 
the \it ROSAT\rm-selected 
Seyfert galaxies. In each figure, the three panels correspond to the three different methods for 
classifying emission-line galaxies using two pairs of line ratios. The galaxies with 
measured $D_n(4000)$ are shown by 
the red open squares, and the galaxies whose spectra are dominated by AGN continuum 
by the blue open circles. At first glance, as compared with the composite AGNs studied in Paper I,
a majority of the \it ROSAT\rm-selected Seyfert galaxies are
located above the theoretical curves discriminating between AGNs and star-forming
galaxies in all the three BPT diagrams (Kewley et al. 2001).
In Figure 4, the size of each point is scaled to be proportional to the value of $D_n(4000)$, if possible. 
All the three panels
indicate a trend that the stellar populations evolve as the Seyfert galaxies deviate from the 
star-forming sequence, which confirms the conclusion that are drawn in Paper I and Kewley et al. (2006) 
by the X-ray selected broad-line AGNs. Figure 5 shows the $D_n(4000)$ index as a function of 
the distance from the starforming sequence for the \it ROSAT\rm-selected Seyfert galaxies.
The distance in each diagnostic diagram is measured from an empirical base point that is marked by a green star 
in Figure 4. All the three panels in Figure 5 show a correlation between $D_n(4000)$ and the 
defined distance\footnote{Similar correlations can be found in the case of H$\delta_A$ instead of 
$D_n(4000)$. The corresponding Spearman rank-order correlation coefficients $r_s$ are -0.471, 
-0.490, -0.623. All the correlations are significant at a level $P<10^{-4}$. }. 
From left to right, Spearman rank-ordered tests yield correlation coefficients 
$r_s=0.379$, 0.367, and 0.482. The corresponding significance levels $P$ are derived to be 0.0006, 0.0021, and 
$<10^{-4}$, where $P$ is the probability of null correlation.
 
In Figure 6, we scale the size of each of the point to be proportional to the value of
$\alpha_X$ instead of $D_n(4000)$. We can not identify a clear trends in the three 
panels as compared with Figure 4. Unlike the cases with $D_n(4000)$, we furthermore 
fail to identify any significant correlation between $\alpha_X$ and the defined distance.
We suspect that the lack of any correlation could be probably caused by the large uncertainty of 
$\alpha_X$ that is resulted 
from the relatively heavy local absorptions in soft X-ray in the partially obscured AGNs.

\subsection{[\ion{O}{1}]/H$\alpha$ vs. $D_n(4000)$ and  [\ion{S}{2}]/H$\alpha$ vs. $D_n(4000)$ correlations}

We found a tight correlation between the line ratio [\ion{O}{1}]/H$\alpha$ and 
$D_n(4000)$ in Paper I for the first time. 
Here, we re-examine the correlation by extending the sample to \it ROSAT\rm-selected Seyfert
galaxies (typically with older stellar populations compared with the composite AGNs, see Section 3.1). 
Figure 7 shows the [\ion{O}{1}]/H$\alpha$ vs. $D_n(4000)$ correlation in the left panel, along with 
the [\ion{S}{2}]/H$\alpha$ vs. $D_n(4000)$ correlation in the right panel. The \it ROSAT\rm-selected 
Seyfert galaxies are shown by the red open circles, and the composite AGNs by 
the green solid circles. The two tight correlations between the two line ratios and $D_n(4000)$ can be clearly 
identified not only for either the \it ROSAT\rm-selected Seyfert galaxies or the composite AGNs, 
but also for the combination of the two samples. Taking the \it ROSAT\rm-selected Seyfert galaxies into account 
only, Spearman rank-ordered tests yield a correlation coefficient $r_s=0.562$ 
at a significance level $P<10^{-4}$ for the [\ion{O}{1}]/H$\alpha$ vs. $D_n(4000)$ correlation,
and $r_s=0.563$ with $P<10^{-4}$ for the [\ion{S}{2}]/H$\alpha$ vs. $D_n(4000)$ correlation.
After combining the two samples, similar tests provide better statistics with the corresponding 
correlation coefficients $r_s=0.661$ ($P<10^{-4}$) and $r_s=0.681$ ($P<10^{-4}$)\footnote{
Similar significant correlations can be still identified for the cases with $H\delta_A$: $r_s=-0.574$, $P<10^{-4}$ for 
[\ion{S}{2}]/H$\alpha$, and $r_s=-0.583$, $P<10^{-4}$ for [\ion{O}{1}]/H$\alpha$.}.
An unweighted least-square fitting results in two calibrations:
\begin{equation}
\log D_{4000}=(0.24\pm0.01)+(0.09\pm0.01)\log([\mathrm{OI}]/\mathrm{H}\alpha)
\end{equation}    
and 
\begin{equation}
\log D_{4000}=(0.20\pm0.01)+(0.14\pm0.01)\log([\mathrm{SII}]/\mathrm{H}\alpha)
\end{equation}
with a value of standard deviation $\sigma=0.04$ for both fittings. The best fitted 
calibrations are shown by the dashed lines in both panels in Figure 7. The 1$\sigma$ deviation
is marked by the dotted lines in each panel. 

\subsection{$L/L_{\mathrm{Edd}}-D_n(4000)$ sequence}

One of the aims of this third paper is to extend the {$L/L_{\mathrm{Edd}}-D_n(4000)$ sequence established 
in Paper I to \it ROSAT\rm-selected Seyfert galaxies. $D_n(4000)$ is plotted against
$L/L_{\mathrm{Edd}}$ in Figure 8 by the red solid squares for the \it ROSAT\rm-selected Seyfert galaxies. 
The composite AGNs studied in Paper I are presented by the blue open circles.  The 
three green open stars mark the position of the three intermediate-$z$ hybrid QSOs studied recently 
in Wang \& Wei (2009). The \it ROSAT\rm-selected Seyfert galaxies follow the evolutionary sequence
that was established previously in the composite AGNs. The correlation coefficient is $r_s=-0.400$ ($P<10^{-4}$)
for the combination of the two samples, although the statistics is poor ($r_s=-0.142, P=0.2464$)
when the \it ROSAT\rm-selected Seyfert galaxies are considered only. 
The poor statistics is most likely caused by the relatively narrow dynamic range 
(see discussion below).

The insert panel in Figure 8 shows the $D_n(4000)-\mathrm{H}\delta_A$ diagram for the \it ROSAT\rm-selected 
Seyfert galaxies. Again, we find a consistence with that reported in Paper I. The mass building in the 
\it ROSAT\rm-selected Seyfert galaxies could be described by a continuous star formation history with 
an exponentially declining star formation rate.

\section{Discussion}

\subsection{Evolutionary significance of $\alpha_X$?}

The sample of the \it ROSAT\rm-selected Seyfert galaxies allows us to examine the evolutionary 
significance of the properties of soft X-ray emission of AGNs. 
We directly plot the two line ratios ([\ion{O}{1}]/H$\alpha$ and [\ion{S}{2}]/H$\alpha$) 
and $D_n(4000)$ as a function of $\alpha_X$ in Figure 9. The top and 
middle panels show two marginal anti-correlations between the two line ratios and 
$\alpha_X$. According to the Spearman rank-ordered tests, the correlation coefficients are
$r_s=-0.274$ with $P=0.0355$ for [\ion{O}{1}]/H$\alpha$ vs. $D_n(4000)$, and $r_s=-0.316$ with $P=0.0171$ for
[\ion{S}{2}]/H$\alpha$ vs. $D_n(4000)$. 
We argue that these dependencies agree with
the theoretical photoionization models. The photoionization calculations in
Kewley et al (2006) suggested that a hard ionizing field with
a power-law index $\alpha<1.4$ is required to produce the strong [\ion{O}{1}] line emission
($\log([\mathrm{OI}]/\mathrm{H}\alpha)\geq-0.6$). 
This scenario seems to be plausible since
various high energy instruments have observed hard X-ray spectra in dozens of LINERs that are
typical of high [\ion{O}{1}]/H$\alpha$ ratios and old stellar populations (e.g., Flohic
et al. 2006; Gliozzi et al. 2008; Rinn et al. 2005).

Given the strong correlation between the two line ratios and $D_n(4000)$, $D_n(4000)$ is 
expected to be related with $\alpha_X$. 
However, no correlation can be identified between $D_n(4000)$ and $\alpha_X$ in the current sample from the 
bottom panel in Figure 9 ($r_s=-0.107, P=0.4184$)\footnote{The Spearman correlation coefficient for the 
correlation between $\alpha_X$ and H$\delta_A$ is $r_s=-0.134, P=0.3122$.}. 
The non-detection of the expected 
correlation might be caused by the uncertainties of both parameters. 
At first, we ignore the local soft X-ray absorption in our estimation of $\alpha_X$.
Secondly, the measurements of
$D_n(4000)$ is model dependent, and highly depends on the sing-to-noise ratios of the spectra at the blue end.
By contrast, the two line ratios can be accurately determined from the observed spectra. 

In summary, we fail to find direct evidence supporting the evolutionary significance of 
$\alpha_X$ from the sample the of \it ROSAT\rm-selected Seyfert galaxies, albeit 
the evolutionary significance could be indirectly supported by the combination of  
the strong [\ion{O}{1}]/H$\alpha$ ([\ion{S}{2}]/H$\alpha$) vs. $D_n(4000)$ correlation 
and identified marginal dependence of 
[\ion{O}{1}]/H$\alpha$ ([\ion{S}{2}]/H$\alpha$) on $\alpha_X$: AGNs with soft X-ray spectrum
might evolve to ones with hard X-ray spectrum as the circumnuclear stellar population continually ages.

\subsection{$L/L_{\mathrm{Edd}}$ vs. $\alpha_X$}

Although the physical origin of the soft excess is still uncertain 
(e.g., Walter \& Fink 1993; Kawaguchi et al. 2001; Schurch \& Done 2007; Turner \& Miller 2009
for a review), it was known for a long time that there is a strong dependence between soft X-ray 
emission and optical spectroscopic properties (and also $L/L_{\mathrm{Edd}}$).
Brandt et al. (1997) found a strong anti-correlation 
between the photon index $\Gamma$ ($N(E)\propto E^{-\Gamma}$, i.e., $\alpha_X=\Gamma-1$) in soft X-ray regime
and FWHM of AGN broad H$\alpha$ emission line (see also in Leighly 1999; Reeves \& 
Turner 2000). Direct correlation analysis and PCA analysis indicated that $\alpha_X$ is
related with the optical spectroscopic properties (e.g., [\ion{O}{3}] emission 
and $\mathrm{RFe= optical\ FeII/H\beta}$) that define the EI space 
(e.g., Grupe 2004; Xu et al. 2003; Laor et al. 1997; Vaughan et al. 1999). Briefly, 
larger the $\alpha_X$, stronger the \ion{Fe}{2} blends and [\ion{O}{3}] emission.

As suggested early in Pounds et al. (1995), a high $L/L_{\mathrm{Edd}}$ state tends to produce a steep
soft X-ray spectrum. 
Thanks to the great progress made in the reverberation mapping technique (e.g., Kaspi et al 2000, 2005; 
Peterson \& Bentze 2006), Boroson (2002) indicated the EI space is mainly physically derived 
by $L/L_{\mathrm{Edd}}$. With these progress in $M_{\mathrm{BH}}$ estimation, 
Grupe (2004) subsequently found a direct correlation between 
$\alpha_X$ and $L/L_{\mathrm{Edd}}$ in a sample of \it ROSAT\rm\ soft X-ray selected type I AGNs.
This correlation was recently confirmed in Desroches et al. (2009) for the AGNs with intermediate $M_{\mathrm{BH}}$.

We fail to identify a correlation between $\alpha_X$
and $L/L_{\mathrm{Edd}}$ ($r_s=-0.127, P=0.3725$) in the current sample, however. In fact,
this is not a surprising result taking into account that both parameters
are of relatively large uncertainties. In addition to the large uncertainties in
the estimation of $\alpha_X$ (see discussion above), the estimation of $L/L_{\mathrm{Edd}}$
is still affected by the orientation effect of AGN.
Paper I stated that the upper limit of the underestimate of $L/L_{\mathrm{Edd}}$
is $\sim50\%$ due to both intrinsic extinction in BLR and torus obscuration.
In fact, Grupe (2004) found a direct correlation between
$\alpha_X$ and $L/L_{\mathrm{Edd}}$ in a sample of \it ROSAT\rm\ soft X-ray selected type I AGNs
whose spectra are predominated by the AGN emission, which was
recently confirmed in Desroches et al. (2009) for AGNs with intermediate $M_{\mathrm{BH}}$.
To avoid the strong absorption in soft X-ray band, we need
to extend our future study into hard X-ray ($\geq$2keV) regime. At present,
the relationship between hard X-ray photon index $\Gamma$ and $L/L_{\mathrm{Edd}}$ is
still a controversial issue (e.g., Shemmer et al. 2008; Constantin et al. 2009).
Current available hard X-ray surveys seem to be strongly biased
towards bright type I AGNs due to
their limited sensitivity (e.g., Wang et al. 2009a).
A deep, well defined hard X-ray survey is therefore necessary in the future for establishing
a better understanding of the evolutionary significance of high energy emission of AGNs.

\subsection{$L/L_{\mathrm{Edd}}-D_n(4000)$ sequence}

There are accumulating observational evidence supporting that $L/L_{\mathrm{Edd}}$ 
plays an important role in the evolution of AGNs. By combining the current sample
with the composite AGNs studied in Paper I, we reinforce 
the $L/L_{\mathrm{Edd}}-D_n(4000)$ sequence
in which $L/L_{\mathrm{Edd}}$ and age of stellar population
are estimated directly from the H$\alpha$ broad emission lines and starlight components, respectively.
Similar relationships were found by other authors 
who used $L_{\mathrm{[OIII]}}/\sigma_*^4$ as a proxy of $L/L_{\mathrm{Edd}}$ (e.g., 
Kewley et al. 2006; Wild et al. 2007; Kauffmann et al. 2007). 
Chen et al. (2009) found a tight correlation between $L_{\mathrm{[OIII]}}/\sigma_*^4$
and specific star formation rate obtained through modeling the continuum and absorption
lines by the single stellar population models. Watabe et al. (2008) evaluated circumnuclear 
starburst luminosity in AGNs by the near-infrared polycyclic aromatic hydrocarbon (PAH) emission, which 
establishes a close correlation between the circumnuclear star formation rate and $L/L_{\mathrm{Edd}}$.

The \it ROSAT\rm-selected Seyfert galaxies studied here are clustered at the middle range of the sequence,
which is likely caused by a selection effect due to the sensitivity limit of RASS
$\sim10^{-13}\ \mathrm{ergs\ s^{-1}\ cm^{-2}}$.
The sensitivity limit therefore leads us to miss some X-ray less-luminous AGNs.
On the one hand, the less-luminous X-ray emission is intrinsic in some AGNs that are
typically associated with relatively old stellar populations and with low $L/L_{\mathrm{Edd}}$
(e.g., Shemmer et al. 2008; Panessa et al. 2006; Hickox et al. 2009). 
On the other hand, there is accumulating evidence that AGNs with both young stellar populations and 
high $L/L_{\mathrm{Edd}}$ tend to be heavily obscured by the plentiful surrounding gas at the beginning of their lives.
The gas is required to not only fuel the central AGN 
activity, but also form circumnuclear stars rapidly. The heavy obscuration causes the AGNs are 
faint in X-ray and optical bands, but typically
luminous in infrared (e.g., Sanders \& Mirabel 1996). Numerical simulations
of merger of two gas rich galaxies including SMBH indicated that the central AGN activities are likely
obscured heavily by the surrounding gas for most time of the starburst 
(e.g., Di Matteo et al. 2005; Springel et al. 2005; Hopkins et al. 2005). Mao et al. (2009)
recently identified three \ion{H}{2}-buried Seyfert galaxies from the SDSS MPA/JHU DR4 catalog
by analyzing their optical spectral properties. Despite the broad Balmer emission lines, the
narrow emission lines of the three Seyfert galaxies are found to be mainly ionized by the power from hot stars.
Daddi et al. (2007) studied the X-ray spectral properties of X-ray undetected IR luminous galaxies at
$z\sim2$ using the X-ray stacking analysis. The complex X-ray analysis suggested that there are
heavily obscured AGNs in these galaxies with intensive star formation activities.

In fact, soft X-ray survey is an efficient way to select narrow-line Seyfert galaxies (NLS1s) that 
are characterized by high $L/L_{\mathrm{Edd}}$, small $M_{\mathrm{BH}}$, and the steepest X-ray spectra 
(e.g., Boroson \& Green 1992; Boroson 2002; Sulentic et al. 2000; Boller et al. 1996; Brandt et al. 1997).
Mathur (2000) argued that NLS1s might be young AGNs that are in the growth phase of their central 
SMBH (see also in e.g., Mathur et al. 2001; Grupe \& Mathur 2004). 
Recent studies found that post-starburst stellar populations are frequently identified in NLS1s 
(e.g., Wang \& Wei 2006; Zhou et al. 2005). In addition to the young stellar populations, 
Sani et al. (2009) recently pointed out that NLS1s are associated with higher ongoing star formation 
activities (using 6.2$\mu$m PAH emission as a tracer) as compared with typical broad-line AGNs.

Comparing the $L/L_{\mathrm{Edd}}-D_n(4000)$ sequence defined by the \it ROSAT\rm-selected 
AGNs with that defined by the optically selected AGNs implies that AGNs with strong X-ray emission represent
a population at a particular evolutionary stage (see also in Hickox et al. 2009). 
The central SMBH is required to be active at the stage 
to produce amount of high energy emission. Meanwhile, in order to detect the X-ray 
emission, the central activity can not be heavily obscured by the surrounding bugle star formation.   
The similar evolutionary scenario was recently proposed in Hickox  et al. (2009) through a 
comprehensive comparison between radio-, X-ray-, and IR-selected AGNs. Compared to the radio-selected 
AGNs with extreme low $L/L_{\mathrm{Edd}}$, the X-ray-selected AGNs have higher $L/L_{\mathrm{Edd}}$. 
Although we can not exclude selection effects totally, Wang et al. (2009a) indicated that 
the RXTE/INTEGRAL hard X-ray selected AGNs have a narrow distribution of 
$L/L_{\mathrm{Edd}}\sim 0.01-0.1$, and low column densities ($<10^{22}\ \mathrm{cm^2}$).  
Moreover, the Galex/SDSS NUV-optical color-magnitude diagram shows that the distribution of X-ray-selected AGNs 
prefer to distribute in 
the ``green valley'' located between the red sequence and the blue cloud (e.g., Shawinski et al. 2009; 
Hickox et al. 2009; Treister et al. 2009).

\section{Summary}

As the third paper of the series, we examine the coevolutionary issue of AGNs and their host 
galaxies by extending the studies in Paper I into \it ROSAT\rm-selected Seyfert 1.8/1.9 galaxies.
These galaxies are selected from the \it ROSAT\rm\ All Sky Survey/SDSS DR5 catalog that was 
originally done by Anderson et al. (2007). Using the similar analysis method adopted in Paper I 
allows us to draw following two main conclusions.

\begin{itemize}
\item Two tight correlations, [\ion{O}{1}]/H$\alpha$ vs. $D_n(4000)$ and [\ion{S}{2}]/H$\alpha$ vs. $D_n(4000)$,
are firmly established not only in the \it ROSAT\rm-selected Seyfert galaxies, but also in the 
combination of the current soft X-ray sample with the composite AGNs studied in Paper I.
The \it ROSAT\rm-selected Seyfert galaxies show that  
the two line ratios depend on the soft X-ray spectra slopes $\alpha_X$ estimated 
from the hardness ratios.

\item The \it ROSAT\rm-selected Seyfert galaxies are well consistent with the 
$L/L_{\mathrm{Edd}}-D_n(4000)$ sequence established in the composite AGNs studied in Paper I.
The \it ROSAT\rm-selected galaxies are, however, not uniformly distributed along the sequence. 
They are clustered at the middle range of the sequence, which could be explained by an 
observational bias. Because of the X-ray count rate threshold of RASS, 
the sample is biased against 
not only the objects with heave absorptions (at the end with high $L/L_{\mathrm{Edd}}$ and 
young stellar population), but also the ones with intrinsic low high energy emission (at the end with low 
$L/L_{\mathrm{Edd}}$ and old stellar population). 

\end{itemize}

%% The displaymath environment will produce the same sort of equation as
%% the equation environment, except that the equation will not be numbered
%% by LaTeX.


%% Included in this acknowledgments section are examples of the
%% AASTeX hypertext markup commands. Use \url without the optional [HREF]
%% argument when you want to print the url directly in the text. Otherwise,
%% use either \url or \anchor, with the HREF as the first argument and the
%% text to be printed in the second.

\acknowledgments

The authors would like to thank the anonymous referee for his/her comments that improve the 
paper. This work was supported by the National Science Foundation 
of China (under grant 10803008) and by the National Basic Research Program of China (grant 2009CB824800). 
The SDSS achieve data are created and distributed by the Alfred P. Sloan Foundation.

\clearpage

%% the file, uncomment the \plotone, \plottwo, and
%% If you need a layout that cannot be achieved with \plotone or
%% \plottwo, you can invoke the graphicx package directly with the
%% \includegraphics command or use \plotfiddle. For more information,
%% please see the tutorial on "Using Electronic Art with AASTeX" in the
%% documentation section at the AASTeX Web site,
%% http://www.journals.uchicago.edu/AAS/AASTeX.
%%
%% The examples below also include sample markup for submission of
%% supplemental electronic materials. As always, be sure to check
%% the instructions to authors for the journal you are submitting to
%% for specific submissions guidelines as they vary from

%% This example uses \plotone to include an EPS file scaled to
%% 80% of its natural size with \epsscale. Its caption
%% has been written to indicate that additional figure parts will be
%% available in the electronic journal.

\begin{figure}
\includegraphics[width = 13cm]{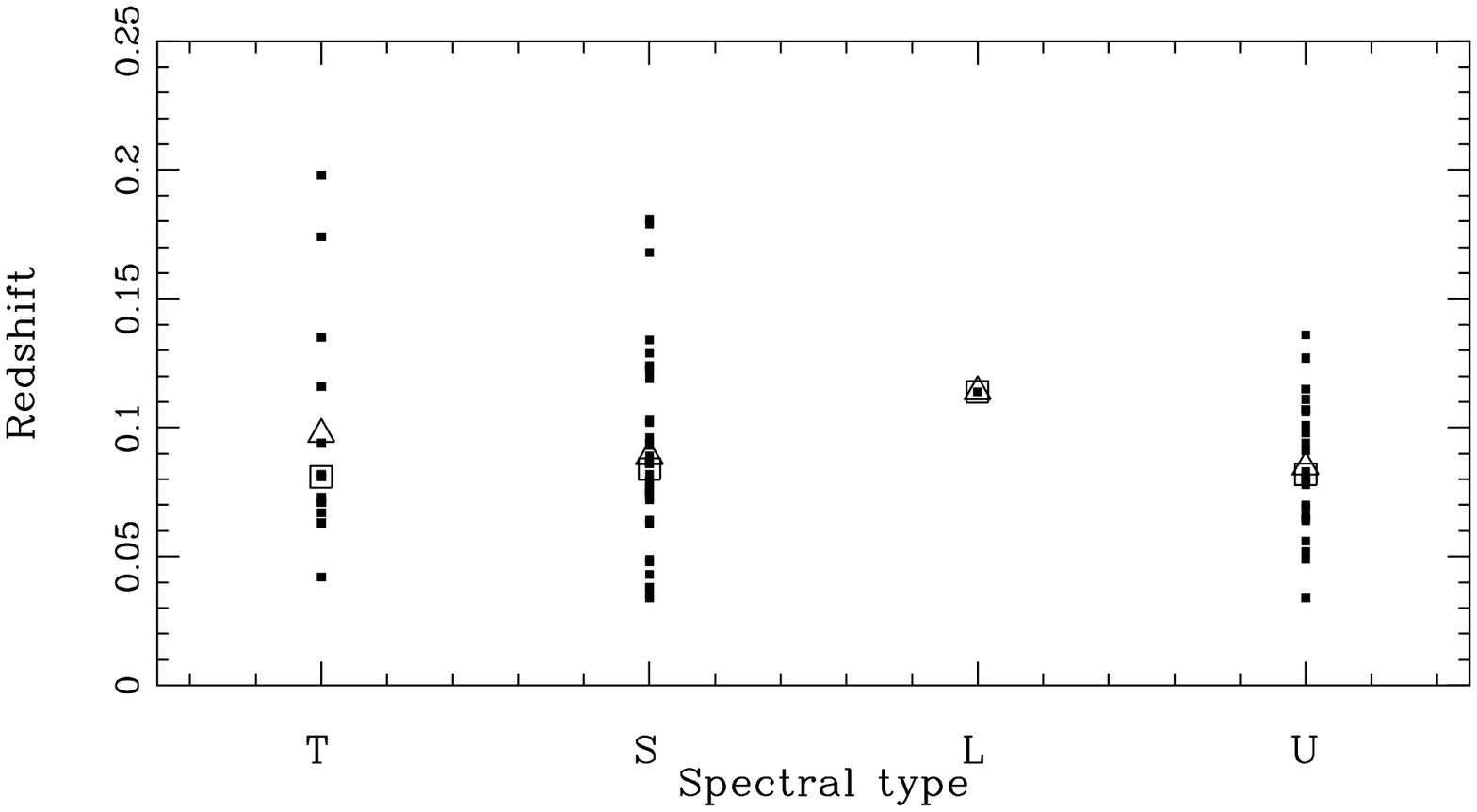}
\caption{Redshifts plotted against the spectral types for the 80 \it ROSAT\rm-selected 
Seyfert galaxies that are used in the current study. All four line ratios are required 
to be measured from the SDSS spectra to perform classification according to the BPT diagrams
(T: transitions, S: Seyfert galaxies, L: LINERs and U: unclassified galaxies).
The mean and median redshifts are marked by the open triangle and square, respectively, for 
each of the spectral type. 
}
\end{figure}

\begin{figure}
\includegraphics[width = 13cm]{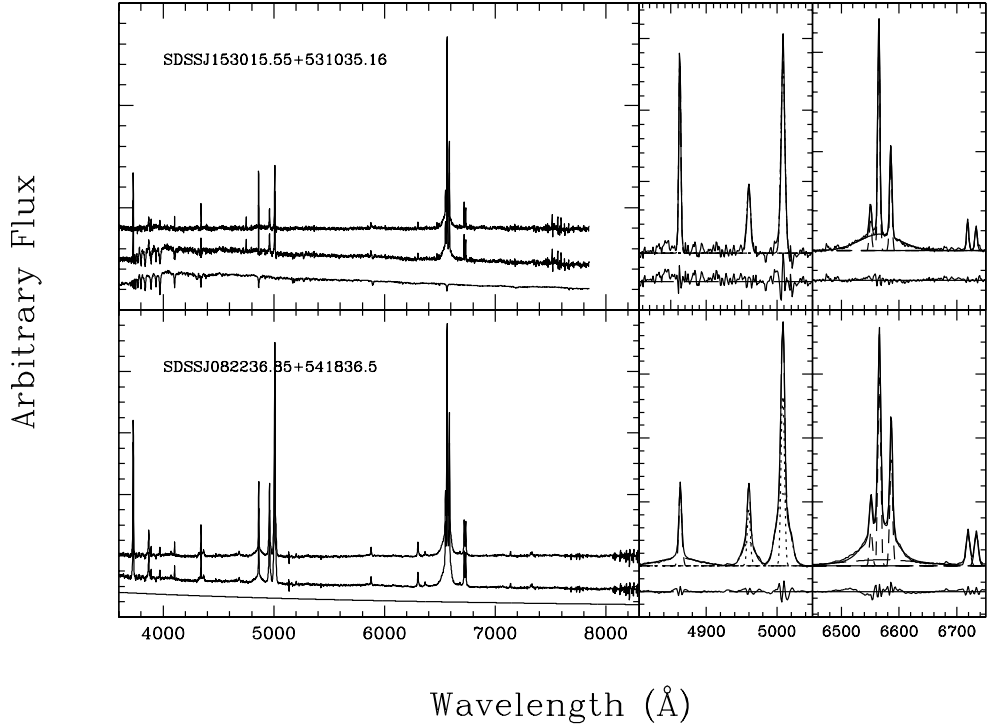}
\caption{\it Left column: \rm the upper panel shows the modeling and subtraction of the starlight components by the
linear combination of the seven reddened eigenspectra. 
We plot the emission-line spectrum, observed spectrum and modeled starlight spectrum
from top to bottom. The spectra are vertically shifted by arbitrary amounts for visibility;
the lower panel is the same as the upper one, but for a case in which the continuum is modeled by a powerlaw. 
\it Middle column: \rm the modelings of the emission-line profiles by a set of Gaussian components 
for H$\beta$ region, The residuals of the profile modelings are displayed under the corresponding spectral
profile in each panel. \it Right column: \rm the same as the middle column but for H$\alpha$ region. 
}
\end{figure}

\clearpage

\begin{figure}
\includegraphics[width = 13cm]{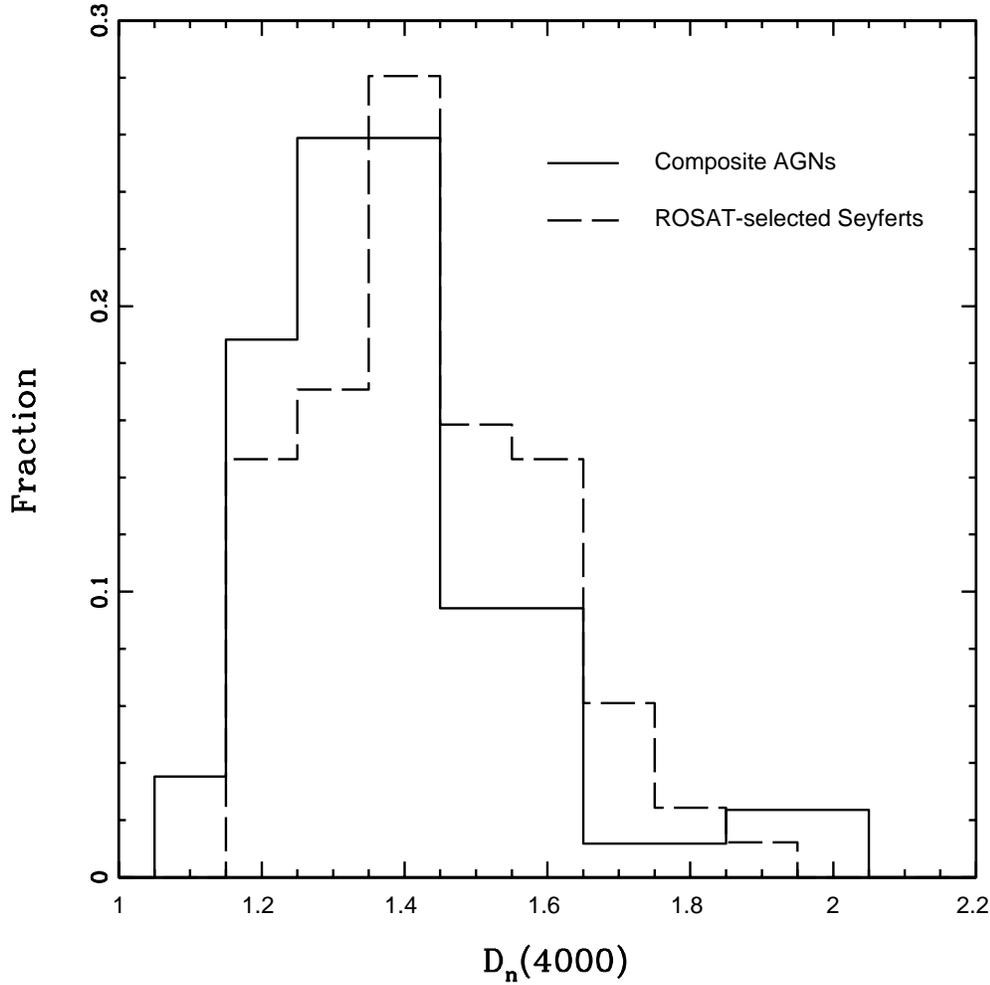}
\caption{Comparison of the index $\mathrm{D_n(4000)}$ between the \it ROSAT\rm-selected Seyfert galaxies and 
the composite AGNs studied in Paper I. The \it ROSAT\rm-selected Seyfert galaxies are plotted by the dashed line, and 
the composite AGNs by the solid line. The comparison indicates that the stellar populations of the 
\it ROSAT\rm-selected Seyfert galaxies are statistically older than that of the composite AGNs. 
}
\end{figure}

%% Here we use \plottwo to present two versions of the same figure,
%% one in black and white for print the other in RGB color
%% for online presentation. Note that the caption indicates
%% that a color version of the figure will be available online.
%%

\begin{figure}
\includegraphics[width = 12cm]{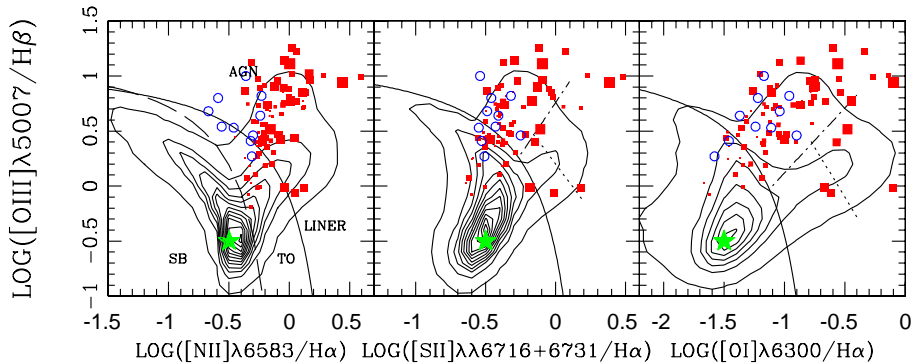}
\caption{The three BPT diagnostic diagrams for the \it ROSAT\rm-selected 
Seyfert galaxies. The red solid points present the galaxies 
whose continuum is dominated by the starlight component, and the 
blue open circles the galaxies whose continuum is dominated by a 
featureless AGN powerlaw. The theoretical demarcation 
lines separating AGNs from star-forming galaxies proposed by Kewley et al. (2001) are shown by the 
solid lines, and the empirical line proposed by Kauffmann et al. (2003a) by the dashed line. The dot-dashed 
lines drawn in the [\ion{S}{2}]/H$\alpha$ vs. [\ion{O}{3}]/H$\beta$ and [\ion{O}{1}]/H$\alpha$ vs. 
[\ion{O}{3}]/H$\beta$ diagrams show the empirical separation scheme of LINERs 
proposed in Kewley et al. (2006). 
The size of each point is proportional to the corresponding value of $D_n(4000)$.
Clear trends with $D_n(4000)$ can be identified in all three panels.  
The stellar populations 
change from young to old as the AGNs deviate from the starforming sequence, which confirms 
the results given in the previous studies (e.g., Wang \& Wei 2008; Kewley et al. 2006). The green star in each 
panel marks the empirical point used to calculate the distance from the starforming sequence.   
}
\end{figure}

\begin{figure}
\includegraphics[width = 12cm]{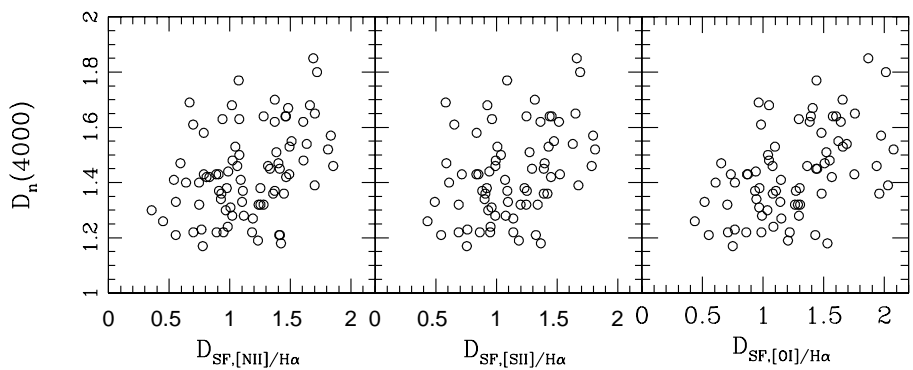}
\caption{The correlations between index $D_n(4000)$ and the distance from the starforming 
sequence (see the definition in the main text) in the three BPT diagrams for the \it ROSAT\rm-selected 
Seyfert galaxies. From left to right, Spearman rank-ordered tests yield correlation coefficients 
$r_s=0.379$, 0.367, and 0.482, and corresponding significance levels $P=0.0006$, 0.0021, and $<10^{-4}$.   
}
\end{figure}

\begin{figure}
\includegraphics[width = 12cm]{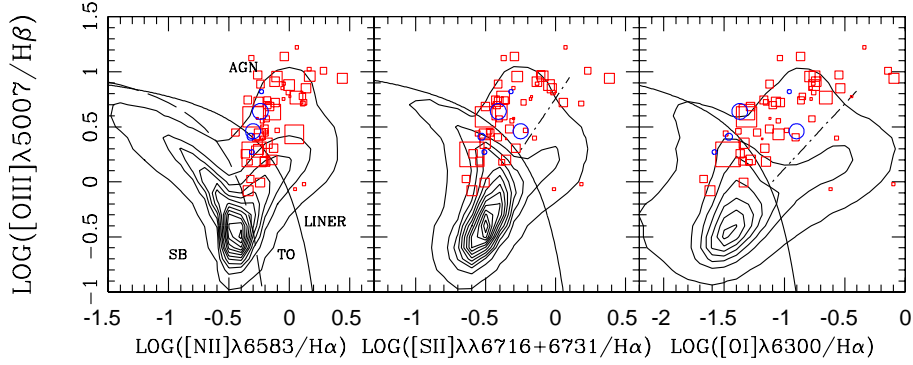}
\caption{The three diagnostic BPT diagrams as same as Figure 4. 
The size of each point is set to be proportional to the corresponding 
value of the soft X-ray spectral slope $\alpha_X$ instead of $D_n(4000)$. 
}
\end{figure}

\begin{figure}
\includegraphics[width = 13cm]{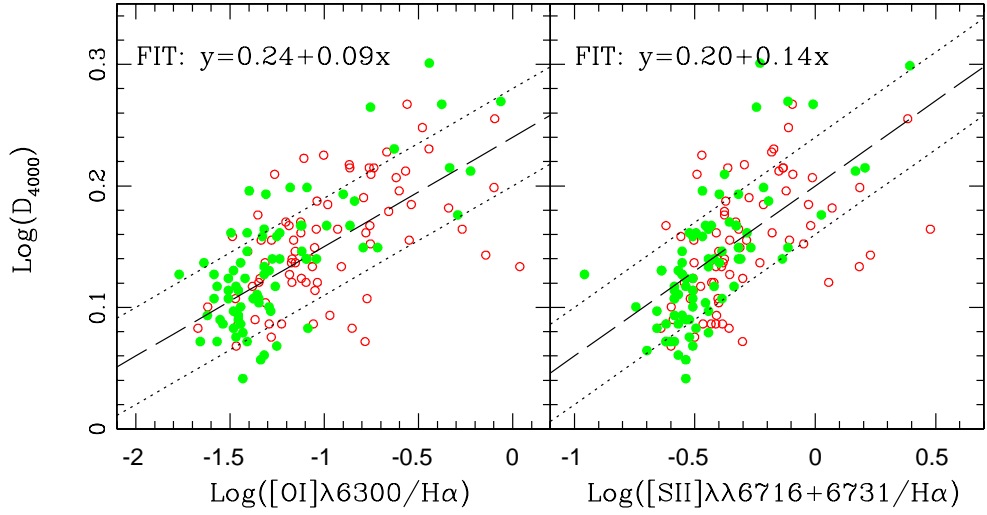}
\caption{The tight correlations [\ion{O}{1}]/H$\alpha$ vs. $D_n(4000)$ (\it left panel\rm) and 
[\ion{S}{2}]/H$\alpha$ vs. $D_n(4000)$ (\it right panel\rm). The \it ROSAT\rm-selected Seyfert 
galaxies are plotted by the red open circles, and the composite AGNs studied in Paper I
by the green solid dots. 
In each panel, the dashed line shows the best fit to the data without 
taking account of the errors of the points. The two dotted lines mark the $1\sigma$ dispersion of the fitting.
}
\end{figure}

\begin{figure}
\includegraphics[width = 13cm]{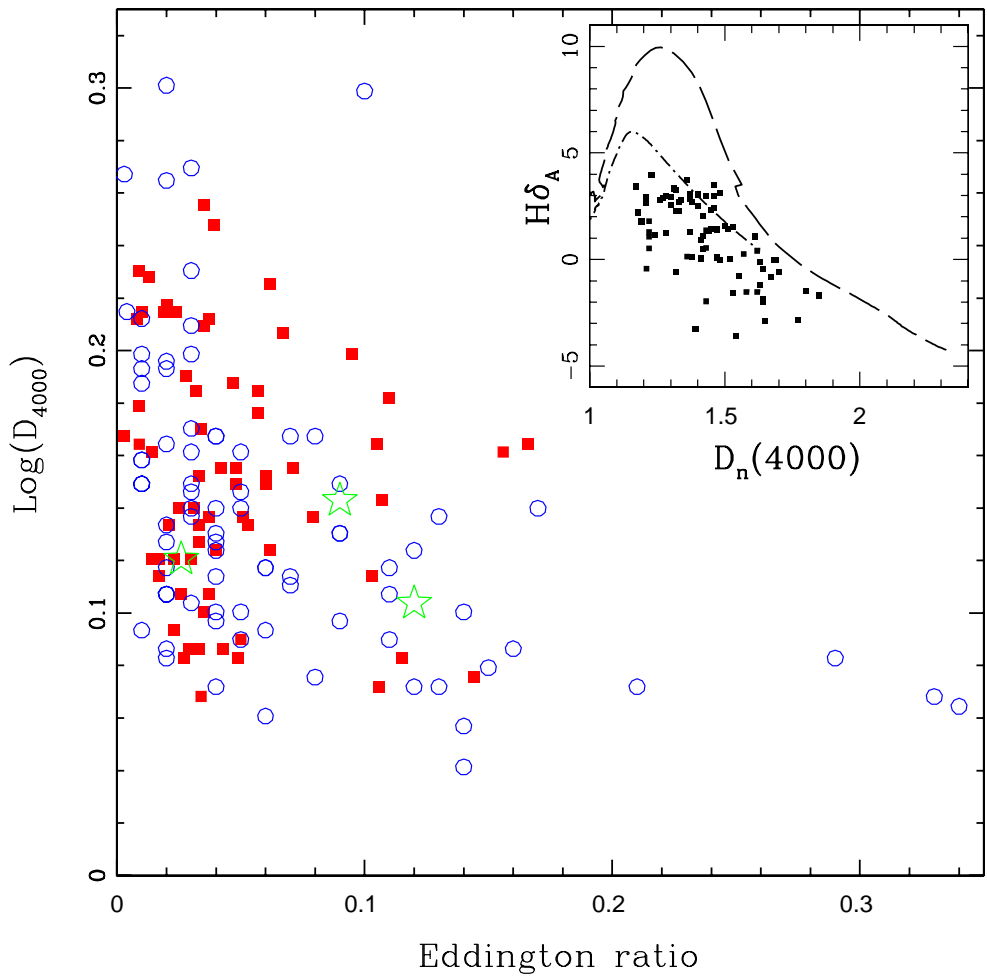}
\caption{$L/L_{\mathrm{Edd}}-D_n(4000)$ sequence. The \it ROSAT\rm-selected Seyfert galaxies 
are plotted by the red solid squares, and the composite AGNs studied in Paper I by the blue open circles. 
The green open stars mark the positions on the diagram for the three intermediate-z AGN-host 
hybrid QSOs studied in Wang \& Wei (2009). All these galaxies follow a common evolutionary trend 
that AGNs co-evolve with their host galaxies in a scenario that $L/L_{\mathrm{Edd}}$ decreases as 
the circumnuclear stellar population continually ages. 
The inserted panel present the $D_n(4000)-\mathrm{H\delta_A}$
diagram for the \it ROSAT\rm-selected Seyfert galaxies. The dashed line shows the stellar 
population evolution locus of the SSP model with solar metallicity, and the dot-dashed line the 
model with exponentially decreasing star formation rate $\Psi(t)\propto e^{-t/(\mathrm{4Gyr})}$. Both models 
are taken from Bruzal \& Charlot (2003). Similar as the composite AGNs, the \it ROSAT\rm-selected 
Seyfert galaxies generally follow the evolutionary locus described by an exponentially decreasing star formation
rate.
}
\end{figure}

\begin{figure}
\includegraphics[width = 13cm]{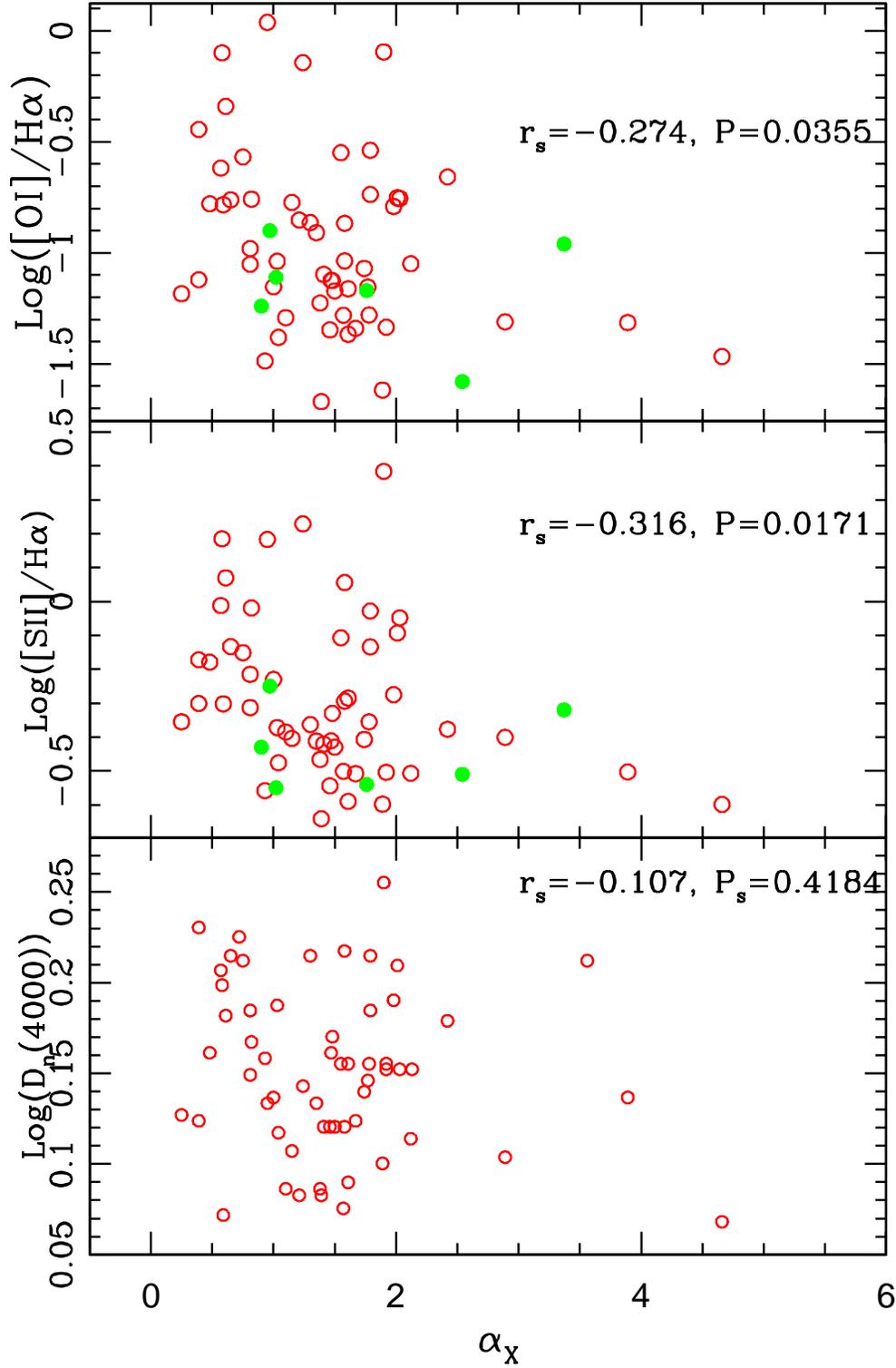}
\caption{Direct dependence of the soft X-ray spectral slope $\alpha_X$ estimated from the 
\it ROSAT \rm hardness ratios
on the two line ratios ([\ion{O}{1}]/H$\alpha$: \it upper panel\rm and [\ion{S}{2}]/H$\alpha$:  
\it lower panel\rm) and on the $D_n(4000)$ index. 
The galaxies with and without determined stellar population ages are shown 
by the red open and green solid circles, respectively. 
}
\end{figure}

%\begin{figure}
%\includegraphics[width = 13cm]{redshift.his.eps}
%\caption{The redshift distribution of the composite AGNs.}
%\end{figure}

%% This figure uses \includegraphics to scale and rotate the still frame

%% If you are not including electonic art with your submission, you may
%% mark up your captions using the \figcaption command. See the
%% No more than seven \figcaption commands are allowed per page,
%% so if you have more than seven captions, insert a \clearpage

%% Tables should be submitted one per page, so put a \clearpage before

%% Two options are available to the author for producing tables:  the
%% deluxetable environment provided by the AASTeX package or the LaTeX
%% table environment.  Use of deluxetable is preferred.
%%

%% Three table samples follow, two marked up in the deluxetable environment,
%% one marked up as a LaTeX table.

%% In this first example, note that the \tabletypesize{}
%% command has been used to reduce the font size of the table.
%% We also use the \rotate command to rotate the table to
%% landscape orientation since it is very wide even at the
%% Note also that the \label command needs to be placed

%% This table also includes a table comment indicating that the full
%% version will be available in machine-readable format in the electronic

%\clearpage

\begin{deluxetable}{ccccccccccccc}
\tabletypesize{\scriptsize}
\rotate
\tablecaption{List of properties of the \it ROSAT\rm-selected Seyfert Galaxies with determined stellar population ages}
\tablewidth{0pt}
\tablehead{
\colhead{SDSS name} & \colhead{z} & \colhead{[\ion{N}{2}]/H$\alpha$} & \colhead{[\ion{S}{2}]/H$\alpha$} &
\colhead{[\ion{O}{1}]/H$\alpha$} & \colhead{[\ion{O}{3}]/H$\beta$} & \colhead{FWHM(H$\alpha$)} & \colhead{L(H$\alpha$)} &
\colhead{$D_n(4000)$} & \colhead{H$\delta_A$} &
\colhead{$L/L_{\mathrm{Edd}}$} & \colhead{$M_{\mathrm{BH}}/M_{\odot}$} & \colhead{$\alpha_X$}
\\
\colhead{} & \colhead{} & \colhead{} & \colhead{} & \colhead{} & \colhead{} & \colhead{$\mathrm{km\ s^{-1}}$} &
\colhead{($10^{42}\ \mathrm{erg\ s^{-1}}$)} & \colhead{} & \colhead{\AA} & \colhead{} & \colhead{} &
\colhead{}\\
\colhead{(1)} & \colhead{(2)} & \colhead{(3)} & \colhead{(4)} & \colhead{(5)} & \colhead{(6)} & 
\colhead{(7)} & \colhead{(8)} & \colhead{(9)} & \colhead{(10)} & \colhead{(11)} & \colhead{(12)} & \colhead{(13)}
}
\startdata
 000202.95-103038.0 & 0.103 & -0.162 & -0.501 & -1.281 & 0.685 & 2342.0 & 2.69 & 1.10 & 1.77 & 0.14 & 7.43 & 1.57\\
 001056.25-090109.9 & 0.081 & -0.244 & -0.383 & -1.174 & 0.375 & 3092.0 & 0.39 & 1.27 & 3.10 & 0.04 & 7.30 & \dotfill\\
 002848.77+145216.3 & 0.089 & -0.299 & -0.503 & -1.314 & 0.590 & 1952.2 & 0.25 & 1.27 & 1.31 & 0.08 & 6.80 & 3.89\\
 003843.06+003451.8 & 0.043 & -0.002 & -0.048 & -0.755 & 0.878 & \dotfill & \dotfill & 1.32 & 2.03 & \dotfill & \dotfill &2.03\\
 010629.30+151449.4 & 0.168 & -0.202 & -0.389 & -1.063 & 0.916 & 5474.1 & 1.78 & 1.27 & 0.13 & 0.02 & 8.11 & \dotfill\\
 011535.54+010839.6 & 0.064 & -0.074 & -0.471 & -1.003 & 0.424 & \dotfill & \dotfill & 1.56 & -0.05 & \dotfill & \dotfill & \dotfill\\
 015610.90+140934.2 & 0.082 & -0.183 & -0.421 & -1.097 & 0.697 & 3802.2 & 0.35 & 1.23 & 3.26 & 0.02 & 7.46 & 1.41\\
 014911.00+130406.9 & 0.073 &  0.122 & -0.230 & -1.153 & 0.722 & 2506.1 & 0.31 & 1.27 & 2.86 & 0.05 & 7.07 & 1.00\\
 032525.35-060837.9 & 0.034 & -0.057 & -0.302 & -0.783 & 0.852 & 2021.9 & 0.61 & 1.09 & 2.19 & 0.11 & 7.01 & 0.59\\
 073646.65+393255.3 & 0.107 &  0.006 & -0.172 & -0.444 & 0.775 & 5112.9 & 0.18 & 1.57 & -0.59 & 0.01 & 7.60 & 0.39\\
 080752.27+383211.0 & 0.067 & -0.242 & -0.508 & -1.340 & -0.008 & 2955.4 & 0.38 & 1.23 & 2.69 & 0.04 & 7.26 & 1.67\\
 082209.55+470552.8 & 0.127 &  0.122 &  0.185 & -0.099 & -0.020 & 1889.2 & 0.33 & 1.46 & -1.53 & 0.10 & 6.83 & 0.58\\
 083552.35+295716.0 & 0.076 & -0.167 & -0.413 & -1.058 & 0.636 & 3951.0 & 0.89 & 1.13 & 1.29 & 0.03 & 7.58 & \dotfill\\
 084257.45+030841.4 & 0.064 & -0.344 & -0.430 & -1.171 & 0.744 & 4117.6 & 0.24 & 1.23 & -0.58 & 0.02 & 7.46 & 1.50\\
 084919.70+531730.0 & 0.111 & -0.201 & -0.476 & -1.381 & 0.460 & \dotfill & \dotfill & 1.21 & 3.33 & \dotfill & \dotfill & 1.04\\
 085824.22+520541.1 & 0.090 & -0.201 & \dotfill & -1.154 & 0.185 & \dotfill & \dotfill & 1.30 & 2.50 & \dotfill & \dotfill & 1.77\\
 085922.63+100132.2 & 0.167 &  0.164 & \dotfill & \dotfill & 1.023 & 2189.5 & 0.25 & 1.55 & -0.02 & 0.06 & 6.91 & 0.72\\
 092740.77+262701.0 & 0.048 & -0.224 & -0.313 & -1.051 & 0.555 & 1660.4 & 0.06 & 1.30 & 0.06 & 0.06 & 6.37 & 0.81\\
 094149.83+611158.8 & 0.123 & -0.189 & -0.404 & -0.773 & 0.572 & 4332.3 & 0.87 & 1.19 & 1.23 & 0.03 & 7.76 & 1.15\\
 094252.55+122443.6 & 0.130 & \dotfill & \dotfill &  -1.204 & 1.021 & 3436.8 & 0.56 & 1.37 & -0.04 & 0.03 & 7.47 & \dotfill\\
 094525.90+352103.5 & 0.208 & -0.005 & -0.358 & \dotfill & 0.818 & 3457.6 & 1.37 & 1.12 & 2.68 & 0.05 & 7.65 & \dotfill\\
 094554.48+423818.6 & 0.075 & -0.136 & -0.401 & -1.311 & 0.634 & \dotfill & \dotfill & 1.18 & 2.90 & \dotfill & \dotfill & 2.89\\
 095623.68+564806.3 & 0.075 & -0.176 & -0.377 & -1.352 & 0.530 & 1510.5 & 0.03 & 1.39 & 1.59 & 0.06 & 6.17 & \dotfill\\
 100151.40+490859.8 & 0.071 & -0.265 & -0.432 & -1.457 & 0.360 & 2706.4 & 0.11 & 1.13 & 0.52 & 0.03 & 6.93 & \dotfill\\
 100634.28+315900.4 & 0.093 & -0.100 & -0.215 & -0.981 & 0.467 & 1642.2 & 0.05 & 1.41 & 1.53 & 0.06 & 6.32 & 0.81\\
 100710.35+113146.1 & 0.082 & -0.207 & -0.544 & -1.347 & 0.188 & 4844.6 & 0.37 & 1.22 & 2.29 & 0.01 & 7.69 & 1.46\\
 100958.85+470943.9 & 0.106 &  0.283 &  0.229 & -0.143 & 1.010 & 1872.5 & 0.42 & 1.29 & -3.25 & 0.11 & 6.87 & 1.24\\
 101234.42+062014.9 & 0.079 & -0.048 & -0.179 & -0.779 & 0.753 & 5969.9 & 1.01 & 1.35 & 1.46 & 0.01 & 8.08 & 0.48\\
 101248.93+555918.5 & 0.126 &  0.104 & \dotfill & -1.108 & 0.853 & \dotfill & \dotfill & 1.55 & -0.81 & \dotfill & \dotfill & \dotfill\\
 101454.23+003420.6 & 0.086 &  0.170 & -0.093 & -0.752 & 0.960 & \dotfill & \dotfill & 1.50 & -1.52 & \dotfill & \dotfill & 2.01\\
 101846.09+345001.6 & 0.035 & -0.266 & -0.301 & -1.121 & 0.577 & 1674.8 & 0.06 & 1.24 & 2.28 & 0.06 & 6.40 & 0.39\\
 101942.53+372828.2 & 0.094 & -0.341 & -0.598 & -1.618 & -0.080 & 2882.2 & 0.24 & 1.17 & 2.81 & 0.04 & 7.15 & 1.89\\
 102039.81+642435.8 & 0.122 & -0.101 & -0.275 & -0.791 & 0.957 & 3303.5 & 0.28 & 1.44 & -0.76 & 0.03 & 7.30 & 1.98\\
 103250.04+545656.8 & 0.119 & -0.205 & -0.330 & -1.125 & 0.480 & \dotfill & \dotfill & 1.37 & 3.12 & \dotfill & \dotfill & 1.48\\
 103519.76+312843.0 & 0.075 &  0.041 & \dotfill & \dotfill & 0.433 & 2208.6 & 0.07 & 1.51 & -1.20 & 0.04 & 6.67 & 3.56\\
 104500.41-001554.0 & 0.091 & -0.262 & -0.107 & -0.549 & 0.972 & 2306.6 & 0.45 & 1.33 & -1.96 & 0.07 & 7.07 & 1.55\\
 111100.55+491942.9 & 0.232 &  0.110 & \dotfill &  -0.852 & 0.775 & 2721.5 & 3.31 & 1.12 & -0.44 & 0.12 & 7.61 & 1.21\\
 111419.24+342843.6 & 0.069 & -0.116 &  0.056 & -1.036 & 0.720 & 3605.4 & 0.51 & 1.23 & 2.29 & 0.03 & 7.49 & 1.58\\
 112852.59-032130.5 & 0.198 & -0.292 & -0.599 & -1.467 & 0.249 & 4880.4 & 3.22 & 1.08 & 3.43 & 0.03 & 8.12 & 4.66\\
 113255.96+051539.6 & 0.101 &  0.440 &  0.384 & -0.095 & 0.941 & 2463.8 & 0.11 & 1.67 & -1.49 & 0.04 & 6.85 & 1.9\\
 115906.90+101001.7 & 0.116 & -0.322 & -0.641 & -1.670 & 0.025 & 4078.9 & 0.73 & 1.12 & 2.91 & 0.03 & 7.67 & 1.39\\
 121600.04+124114.3 & 0.066 & -0.243 & -0.507 & -1.049 & 0.432 & 4690.2 & 0.46 & 1.20 & 2.91 & 0.02 & 7.71 & 2.12\\
 122131.04+000512.8 & 0.107 & -0.446 & -0.466 & -1.226 & 0.447 & 3531.9 & 0.58 & 1.13 & 1.80 & 0.03 & 7.50 & 1.38\\
 122602.46+004640.0 & 0.083 & -0.233 & -0.410 & -0.970 & 0.447 & 3322.9 & 0.18 & 1.15 & 1.12 & 0.02 & 7.21 & \dotfill\\
 123155.13+323240.2 & 0.065 & -0.278 & -0.412 & -0.909 & 0.404 & 2654.3 & 0.14 & 1.26 & 3.73 & 0.03 & 6.97 & 1.35\\
 123942.53+342456.3 & 0.098 & -0.141 & -0.355 & -1.280 & 0.200 & 2827.8 & 0.34 & 1.32 & 1.37 & 0.04 & 7.20 & 1.78\\
 124307.02+421231.0 & 0.072 & -0.254 & -0.407 & -1.070 & 0.728 & 2553.5 & 0.10 & 1.28 & 0.12 & 0.03 & 6.87 & 1.74\\
 125055.53-021345.4 & 0.086 & -0.078 & -0.318 & -1.042 & 0.475 & 4524.9 & 0.10 & 1.36 & 2.42 & 0.01 & 7.37 & \dotfill\\
 131235.16+494224.6 & 0.129 &  0.025 & -0.121 & -0.602 & 1.257 & \dotfill & \dotfill & 1.46 & 0.26 & \dotfill & \dotfill & \dotfill\\
 132047.79+320720.4 & 0.071 & -0.282 & -0.401 & -1.406 & 0.101 & \dotfill & \dotfill & 1.30 & 3.04 & \dotfill & \dotfill & \dotfill\\
 132201.77-001814.4 & 0.075 & -0.182 & -0.363 & -0.864 & 0.740 & 3003.4 & 0.13 & 1.52 & -1.86 & 0.02 & 7.05 & 1.30\\
 134243.62+050432.2 & 0.136 &  0.137 & -0.028 & -0.538 & 0.850 & 2979.3 & 0.24 & 1.41 & -1.58 & 0.03 & 7.17 & 1.79\\
 134419.60+512624.6 & 0.063 & -0.121 & -0.134 & -0.737 & 0.909 & 5537.1 & 0.31 & 1.52 & -0.43 & 0.01 & 7.78 & 1.79\\
 135244.63-001526.2 & 0.124 &  0.059 & -0.377 & -0.658 & 0.767 & 6331.4 & 0.46 & 1.40 & 1.46 & 0.01 & 7.98 & 2.42\\
 140844.49+535128.4 & 0.083 & -0.255 & \dotfill & \dotfill & 0.296 & 1857.0 & 0.10 & 1.31 & 1.11 & 0.06 & 6.58 & 1.92\\
 141231.64+435536.1 & 0.094 & -0.252 & -0.519 & -1.471 & 0.490 & 2604.6 & 0.17 & 1.18 & 2.97 & 0.04 & 6.99 & \dotfill\\
 141611.95+631128.0 & 0.077 & -0.364 & -0.491 & -1.263 & 0.864 & 3564.7 & 0.71 & 1.50 & 0.41 & 0.04 & 7.55 & \dotfill\\
 143937.48+514045.9 & 0.078 & -0.194 & -0.558 & -1.487 & 0.439 & \dotfill & \dotfill & 1.34 & 1.34 & \dotfill & \dotfill & 0.93\\
 144242.62+011911.0 & 0.034 & -0.041 & -0.181 & -0.668 & -0.014 & 2736.6 & 0.02 & 1.56 & -0.05 & 0.01 & 6.57 & \dotfill\\
 144248.94+550910.7 & 0.107 &  0.002 & -0.095 & -0.560 & 1.113 & \dotfill & \dotfill & 1.71 & -1.68 & \dotfill & \dotfill & \dotfill\\
 144750.45+051752.0 & 0.086 & -0.041 & -0.294 & -0.867 & 1.139 & 4232.6 & 0.43 & 1.53 & -2.90 & 0.02 & 7.60 & 1.58\\
 144921.57+631613.9 & 0.042 &  0.597 &  0.478 & -0.268 & 0.994 & 1607.6 & 0.19 & 1.35 & 3.48 & 0.11 & 6.57 & \dotfill\\
 150801.34+462200.6 & 0.094 & -0.314 & -0.372 & -1.038 & 1.125 & 2807.3 & 0.43 & 1.43 & -3.61 & 0.05 & 7.23 & 1.03\\
 151640.21+001501.8 & 0.052 &  0.112 &  0.183 & 0.038  & 0.712 & 2274.9 & 0.20 & 1.26 & -2.38 & 0.05 & 6.90 & 0.95\\
 152158.52+540151.5 & 0.056 & -0.243 & -0.355 & -1.184 & 0.390 & 2239.4 & 0.06 & 1.25 & 2.80 & 0.03 & 6.64 & 0.25\\
 153015.55+531035.1 & 0.174 & -0.364 & -0.590 & -1.367 & 0.255 & 3789.0 & 2.34 & 1.14 & 3.97 & 0.05 & 7.84 & 1.61\\
 153446.62+423536.8 & 0.073 & -0.329 & -0.385 & -1.292 & 0.178 & 2764.5 & 0.33 & 1.13 & 1.08 & 0.04 & 7.17 & 1.1\\
 153822.79+551618.8 & 0.165 & -0.053 & \dotfill & -0.929 & 0.738 & 1778.8 & 0.95 & 1.35 & 2.97 & 0.17 & 6.98 & \dotfill\\
 154724.51+001938.8 & 0.212 & -0.267 & \dotfill & \dotfill & 0.274 & 5395.3 & 4.87 & 1.32 & 0.51 & 0.03 & 8.29 & 2.13\\
 162039.49+260305.0 & 0.096 & -0.129 & -0.285 & -1.162 & 0.307 & \dotfill & \dotfill & 1.32 & 2.97 & \dotfill & \dotfill & 1.61\\
 163348.87+355318.5 & 0.038 & -0.007 & -0.019 & -0.758 & 0.812 & 7155.0 & 0.06 & 1.37 & 0.11 & 0.003 & 7.69 & 0.82\\
 164430.26+245728.3 & 0.102 & -0.238 & -0.411 & -1.124 & 0.888 & 2093.6 & 1.87 & 1.34 & 2.30 & 0.16 & 7.26 & 1.47\\
 164520.62+424528.0 & 0.049 & -0.222 & -0.151 & -0.568 & 0.398 & 4116.1 & 0.04 & 1.51 & -0.12 & 0.01 & 7.12 & 0.75\\
 171009.86+642256.6 & 0.256 & -0.320 & \dotfill & \dotfill & -0.193 & 2619.8 & 2.09 & 1.20 & 2.55 & 0.10 & 7.48 & \dotfill\\
 171414.70+650626.1 & 0.079 & -0.046 & -0.133 & -0.761 & 0.892 & 3881.2 & 0.26 & 1.52 & -1.99 & 0.02 & 7.42 & 0.65\\
 171657.25+643312.9 & 0.081 & -0.353 & -0.620 & -1.187 & 0.075 & \dotfill & \dotfill & 1.37 & 1.42 & \dotfill & \dotfill & \dotfill\\
 171811.93+563956.1 & 0.114 &  0.047 & -0.012 & -0.618 & -0.066 & 1760.2 & 0.10 & 1.5 & 1.06 & 0.07 & 6.53 & 0.57\\
 172350.14+653404.2 & 0.170 & -0.264 & \dotfill & \dotfill & -0.015 & 2898.2 & 0.53 & 1.31 & 0.93 & 0.05 & 7.30 & \dotfill\\
 172643.83+600238.7 & 0.080 & -0.154 & -0.377 & -1.170 & 0.412 & 3085.3 & 0.15 & 1.28 & 2.70 & 0.03 & 7.12 & \dotfill\\
 173207.20+542640.5 & 0.115 & -0.183 & -0.504 & -1.335 & 0.351 & 2817.9 & 0.46 & 1.33 & 0.54 & 0.05 & 7.25 & 1.92\\
 222852.76-090452.5 & 0.070 & -0.153 & -0.111 & -0.479 & 0.516 & 3446.5 & 0.79 & 1.64 & -2.85 & 0.03 & 7.54 & \dotfill\\
 230417.31-081646.3 & 0.080 &  0.061 &  0.070 & -0.340 & 1.222 & 1743.1 & 0.32 & 1.41 & 0.04 & 0.11 & 6.75 & 0.61\\
\enddata
%% Text for table notes should follow after the \enddata but before
%% the \end{deluxetable}. Make sure there is at least one \tablenotemark
%% in the table for each \tablenotetext.
\end{deluxetable}

%% If you use the table environment, please indicate horizontal rules using
%% Do not put multiple tabular environments within a single table.
%% The optional \label should appear inside the \caption command.

\clearpage
\begin{deluxetable}{ccccccc}
\tabletypesize{\footnotesize}
\tablecaption{List of properties of the \it ROSAT\rm-selected Seyfert Galaxies whose continuum is dominated 
by AGN powerlaw emission}
\tablewidth{0pt}
\tablehead{
\colhead{SDSS name} & \colhead{z} & \colhead{[\ion{N}{2}]/H$\alpha$} & \colhead{[\ion{S}{2}]/H$\alpha$} &
\colhead{[\ion{O}{1}]/H$\alpha$} & \colhead{[\ion{O}{3}]/H$\beta$} & \colhead{$\alpha_X$}\\
\colhead{(1)} & \colhead{(2)} & \colhead{(3)} & \colhead{(4)} & \colhead{(5)} & \colhead{(6)} &
\colhead{(7)}\\ 
}
\startdata
113520.76+632042.5 & 0.179 & -0.36 & -0.54 & -1.17 & 1.00 & 1.76\\ 
004551.46+155547.6 & 0.115 & -0.56 & -0.43 & -1.24 & 0.54 & 0.90\\ 
011932.86+000837.2 & 0.181 & -0.24 & -0.41 & -1.37 & 0.64 & \dotfill\\ 
082236.85+541836.5 & 0.086 & -0.23 & -0.32 & -0.96 & 0.82 & 3.37\\ 
110157.90+101739.3 & 0.034 & -0.46 & -0.55 & -1.11 & 0.53 & 1.02\\ 
121651.75+375438.0 & 0.063 & -0.32 & -0.53 & -1.46 & 0.41 & \dotfill\\ 
123429.85+621806.2 & 0.135 & -0.31 & -0.51 & -1.58 & 0.27 & 2.54\\ 
142119.12+631300.1 & 0.134 & -0.67 & -0.49 & -1.04 & 0.68 & \dotfill\\ 
162952.88+242638.3 & 0.038 & -0.30 & -0.25 & -0.90 & 0.46 & 0.97\\ 
165601.60+211241.1 & 0.049 & -0.59 & -0.46 & -1.22 & 0.80 & \dotfill\\
\enddata
%% Text for table notes should follow after the \enddata but before
%% the \end{deluxetable}. Make sure there is at least one \tablenotemark
%% in the table for each \tablenotetext.
\end{deluxetable}

%% from page to page when the default \tablewidth is used, as below.  The
%% individual table widths for each page will be written to the log file; a
%% maximum tablewidth for the table can be computed from these values.
%% The \tablewidth argument can then be reset and the file reprocessed, so
%% that the table is of uniform width throughout. Try getting the widths
%% from the log file and changing the \tablewidth parameter to see how
%% adjusting this value affects table formatting.

%% The \dataset{} macro has also been applied to a few of the objects to
%% show how many observations can be tagged in a table.

\clearpage

%% Tables may also be prepared as separate files. See the accompanying
%% sample file table.tex for an example of an external table file.
%% To include an external file in your main document, use the \input
%% command. Uncomment the line below to include table.tex in this
%% sample file. (Note that you will need to comment out the \documentclass,
%% \begin{document}, and \end{document} commands from table.tex if you want
%% to include it in this document.)


%% The following command ends your manuscript. LaTeX will ignore any text

\end{document}